\documentclass[12pt]{amsart}
\usepackage{amsmath,amssymb, amsfonts,amscd,graphicx,stmaryrd,natbib, mathabx}
\usepackage{epsfig}
\usepackage{epstopdf}
\usepackage{subfigure}
\usepackage{amsthm}
\usepackage{fullpage}
\usepackage{verbatim}
\usepackage{listings}
\usepackage{color}
\usepackage{mathtools}
\usepackage{multicol}

\newcommand{\lebn}

\usepackage[misc]{ifsym}
\usepackage{caption}
\usepackage{float}

\usepackage{subfig}

\theoremstyle{plain}
\newtheorem{prop}[equation]{Proposition}

\newtheorem{thm}[equation]{Theorem}

\theoremstyle{definition}

\newtheorem{defn}[equation]{Definition}

\numberwithin{equation}{section}

\begin{document}

\bibliographystyle{plain}

\title[Wolbachia]{An algebraic discussion of bisexual populations with Wolbachia infection I: Discrete Dynamical System Approach}
\author{Barış ÖZDİNÇ, Songül ESİN, Müge KANUNİ}
\address{CosmosID, Suite 300, 20030 Century Blvd, Germantown, MD 20874 U.S.A.}
\email{baris@cosmosid.com }
\address{Department of Mathematics and Computer Science, İstanbul Kültür University, Ataköy
Kampüsü, Bakırköy 34158, İstanbul, Turkey.}
\email{s.esin@iku.edu.tr}
\address{Department of Mathematics, Düzce University,
Düzce, 81620, Turkey.}
\email{mugekanuni@duzce.edu.tr}


\date{\today}
\thanks{The authors would like to thank Can Er for writing the code using the Python programming language and generating the graphs of the model in the Computations and Data Analysis Section of the text.}
\begin{abstract}
This is the first paper in the sequel studying the Wolbachia-infection in bisexual populations. This paper considers the behavior of the population as a discrete dynamical system. The recurrence relation is obtained as a function of the initial infected male/female frequencies and the cytoplasmic incompatibility of the population. The experimental data from Wolbachia-infected terrestrial isopod populations and the model proposed in Wolbachia-infected mosquitoes from literature is compared with the discrete dynamical system achieved.   
\end{abstract}
\maketitle
\section{Introduction}  \label{intro}
Wolbachia is a sexual parasite that exploits the sexual reproduction of its host for itself. It  is widespread in insects and can be transferred to gametes, which may lead to the feminization of the host embryo, and even embryonic mortality through male killing or cytoplasmic incompatibility.  

The widespread prevalence of Wolbachia in almost 40 percent of arthropod species \citep{Zug}, as well as the potential use of Wolbachia sexual manipulation of the host to control insect reproduction dynamics, led to studies investigating Wolbachia infection dynamics in insects such as honeybees ({\it Apis mellifera carnica}) in Germany \citep{WolbachBee}; isopods from France \citep{isopod}, or male Wolbachia-infected mosquitoes to reduce the dengue disease incidences in Singapore \citep{Mos-Sing}. On the one hand, the study from Singapore illustrates that Wolbachia trans-infected male introduction to a dengue-infected mosquito community reduces the vector competency of dengue and cripples dengue transmission. On the other hand, Wolbachia emerges to induce cytoplasmic incompatibility (CI) in insect species such as bees, isopods, and mosquitoes. CI is the phenotypic expression of Wolbachia-infected hosts where eggs from the uninfected female and Wolbachia-infected male are unviable. Therefore, Wolbachia infection may cripple the reproduction and fitness of insect communities that could be pests or disease vectors.

Wolbachia infection of CI-inducing strains often leads to fixation in natural host populations driving reproductive isolation within species. However, empirical evidence suggests different parameters affecting Wolbachia fixation dynamics in host populations, such as initial Wolbachia frequency, host competition, or Wolbachia CI intensity. Therefore, quantifying CI-inducing Wolbachia frequency dynamics would be founding a better understanding of how natural populations of insects could be controlled in the wild. 

In \citep{Fine}, the author models the Wolbachia-infection of mosquitoes as a discrete dynamical system with some parameters such as maternal vertical transmission rate (maternal CI), paternal gamete affection rate (paternal CI), fertility rate, relative survival rate, the prevalence rate of Wolbachia infection in the population. The Equation \ref{recrel} of this paper is a special case of \citep[Equation 1]{Fine}. In this study, the fertility rate and the survival rate are considered to be 1. 

In this sequel of articles, the aim is to discuss the Wolbachia-infected population from two different mathematical aspects. One approach will consider the infected population via a discrete-time dynamical system which is the scope of this article numbered as I, whereas the other approach will consider the population as an evolution algebra and will focus on the algebraic properties similar and different to the evolution algebra of a bisexual population (See \cite{II}). The mathematical theories of both methods are already established, hence Wolbachia-infected population will be a particular example to implement the results known and characterize its algebraic properties.

In this first paper, the discussion is on the discrete dynamical system approach to the Wolbachia-infected population. The outlay of the paper will be as follows: Section \ref{section:prel} starts with the preliminaries from biology and dynamical systems. Section \ref{section:dynamical} will consider the Wolbachia-infected population as a discrete dynamical system with a given cytoplasmic incompatibility and the stable/unstable equilibrium points of the system are calculated. As a deviation from \citep{Fine}, and \citep{Turelli} the non-equal male and female Wolchaia-infection frequencies are introduced to the system. Proposition \ref{xequalsy} shows that even if the frequencies of male and female Wolbachia-infected populations differ initially, it takes only one generation for the frequencies to equal. 

Section \ref{section:Codes} analyzes the experimental data from the literature with the dynamical system model constructed (Equation \ref{recrel}) in Section \ref{section:dynamical}, using a code in Python (which is posted at https://github.com/canbluebird/Wolbachia). Hence, the theory is simulated for 100+ generations, and Wolbachia-infected population frequencies are graphed to reveal the equilibrium points of the system with different cytoplasmic incompatibility (CI) values and initial infected population rates of males and females. The theoretical results are compared with the experimental data of the Wolbachia-infected terrestrial isopod population in \citep{isopod} and the threshold value of the Wolbachia-infected mosquito populations in \citep{Wolbachfigure}. 

\section{Preliminaries}\label{section:prel}
\subsection{Biology}
We start with some basic terminology from biology.
\begin{defn}\cite{EcoEvo}\label{defn:CI}
 \emph{Cytoplasmic incompatibility (CI)} is the reproductive incompatibility
between males infected with a particular strain of bacteria and females not infected with this strain. 
\end{defn}

\begin{figure}[ht]
\begin{center}
\includegraphics[scale=0.30]{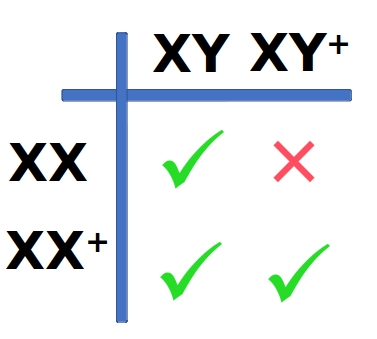}
\end{center}
\caption{Crosses involving the cytoplasmic incompatibility.}
\label{fig:bm-CI-Table} 
\end{figure}

Offspring from an infected male and uninfected female suffer up to 100\% embryonic mortality. In Figure \ref{fig:bm-CI-Table}, the cross represents that the offspring dies and the check represents that the offspring survives from the cross of the infected (marked with +) and uninfected females and males.    

The figure basically gives a crossing table of the phenotype of the population. On the other hand, the infected female/male produces gametes that are both infected and uninfected. 

\begin{table}[H]
\scalebox{0.8}{
\begin{tabular}{|c|c|c|c|c|}
\hline
\multicolumn{1}{|c|}{{\small Crossing}}  & $X$ & $X^{+}$ & $Y$ & $Y^{+}$ \\ \hline
\multicolumn{1}{|c|}{$%
\begin{array}{c}
\text{ } \\ 
\text{ }%
\end{array}%
X%
\begin{array}{c}
\text{ } \\ 
\text{ }%
\end{array}%
$} & uninfected female & (death) no offspring & uninfected male
& (death) no offspring \\ \hline
\multicolumn{1}{|c|}{$%
\begin{array}{c}
\text{ } \\ 
\text{ }%
\end{array}%
X^{+}%
\begin{array}{c}
\text{ } \\ 
\text{ }%
\end{array}%
$} & infected female & infected female & infected male & 
infected male\\ \hline
\end{tabular} } 
\caption{\small Crossing of the gametes with/without Wolbachia infection.}
\end{table}
 The cytoplasmic incompatibility of the population is given as $p$ where $p$ is a real number between 0 and 1, hence $100p$ is the percentage of the transmission of the Wolbachia infection from an individual onto its offspring. 
Some literature considers the cytoplasmic incompatibility among genders as two separate parameters. For instance, in \citep{Fine} maternal vertical transmission rate $d$ and male gamete affection rate $w$ are used as independent variables. However, in this paper, $p=d=w$ is considered. 
 If $p=1$, the infected individual produces all infected gametes, if $p=0$, then the infected individual produces no infected gametes. 

\subsection{Dynamical system}

We need some preliminaries that we quote from \citep{Galor, Rozikov-billards, MathInsight}. 

A dynamical system models the evolution of some quantities over time. If this evolution occurs smoothly over time, then the system is {\it continuous-time}, if evolution occurs in discrete time steps, then it is a {\it discrete dynamical system}. 
 In such a model, we determine the variables that will evolve over time and the rule that specifies how that variable evolves with time. 
 
 The variables are called the state variables. The set of all the possible values of the state variables is the state space. The state space can be discrete, consisting of isolated points, such as if the state variables could only take on integer values. It could be continuous, consisting of a smooth set of points, such as if the state variables could take on any real value. In the case where the state space is continuous and finite-dimensional, it is often called the phase space, and the number of state variables is the dimension of the dynamical system. The state space can also be infinite-dimensional. We denote the state space by $X$.

  The time evolution rule could involve discrete or continuous time. If the time is discrete, then the system evolves in time steps, and we usually let the time points be the integers $t=0,1,2, \dots$. We can write the state of the system at time $t$ as $x_t \in X$, hence the time evolution rule will be based on a function $f$ that takes as its input the state of the system at one time and gives as its output the state of the system at the next time. Starting at the initial conditions $x_0$ at time $t=0$, we can apply the function $f: X \rightarrow X$ once to determine the state $x_1=f(x_0)$ at time $t=1$, apply the function a second time to get the state $x_2=f(x_1)$ at time $t=2$, and continue repeatedly applying the function to determine all future states. We end up with a sequence of states, the trajectory of the point $x_0: x_1,x_2,x_3, \dots$. In this way, the state at all times is determined both by the function $f$ and the initial state $x_0$ (\citep{MathInsight}). We refer to such as system as a {\it discrete dynamical system.}
 

A point $x^*$ is called a {\em fixed point (equilibrium)} of the dynamical system, if $f(x^*)=x^*$. In a dynamical system, the main concern is to find and classify fixed points. Hence, consider a dynamical system where $X$ is a subset of $\mathbb{R}$ and $f$ is a continuous and differentiable function on $X$, a fixed point $x^* \in X$ is called {\em hyperbolic} if $|f'(x^*)|\neq 1$ (\citep[Definition 1.5]{Rozikov-billards}).

The following theorem provides the main tool to check the type of the ﬁxed point. 
\begin{thm}\citep[Theorem 1.1]{Rozikov-billards} \label{thm:fixedpt} Let $X \subset R$ and $f$ be continuously differentiable on $X$. Let $x \in X$ be a hyperbolic ﬁxed point of $f$ then 
\begin{itemize}
\item[(i)] If $|f'(x^*)| < 1$, then $x^*$ is attracting (stable).
\item[(ii)] If $|f'(x^*)| > 1$, then $x^*$ is repelling (unstable).
\end{itemize}
\end{thm}

Given a discrete dynamical system, the question is to find the fixed points of the system. (i.e. the equilibria). 
The major techniques used for finding the solution to the problem consist of a graphical approach and an analytical approach.  

\begin{figure}[H]
\begin{center}
\includegraphics[scale=0.5]{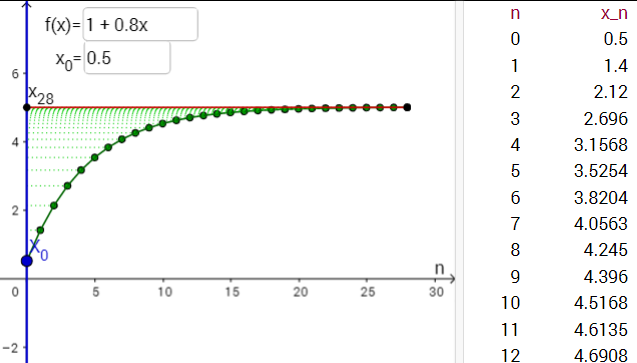}
\end{center}
\caption{A dynamical system model with a stable fixed point. \\
{\it Produced by the Applet: Function iteration from \cite{MathInsight}} }
\label{Applet-MathInsight} 
\end{figure}

Figure \ref{Applet-MathInsight} illustrates a dynamical system model of $f(x)=1+0.8x$ with initial state $x_0= 0.5$. The derivative of $f$ is 0.8 for all values of $x \in X$ and the analytical solution to the equation $f(x^*)=x^*$ gives the fixed point $x^*=5$. Hence, the system has an equilibrium at $5$ and this is a stable fixed point. On the graphical approach, the figure \ref{Applet-MathInsight} shows $n=28$ iterations and the value of the function as the right column and supplies the graph of the data on the left.  

\begin{figure}[h]
\begin{center}
\includegraphics[scale=0.5]{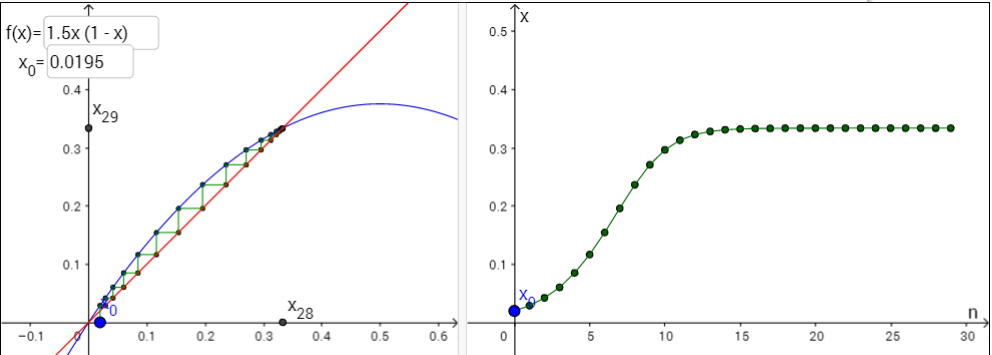}
\end{center}
\caption{A dynamical system model illustrating an unstable fixed point. \\
{\it Produced by the applet: Visualizing function iteration via cobwebbing in \cite{MathInsight}} }
\label{Unstable1-Applet} 
\end{figure}

Now, consider another example of a dynamical system with $f(x)=1.5x(1-x)$. Here, $f$ contains two fixed points $0$ and $\frac{1}{3}$. In this system, fixed points $0$ and $\frac{1}{3}$ are found analytically by solving the equation $f(x^*)=x^*$. Since $f'(x)=1.5-3x$, $f'(0)=1.5 > 1$ and $f'(\frac{1}{3})=0.5 < 1$. By Theorem \ref{thm:fixedpt}, $x^*=0$ is an unstable fixed point whereas $x^*=\frac{1}{3}$ is a stable fixed point. If the system is initially at $0$, it remains at the equilibrium $0$, however when the initial state is slightly changed to say $x_0= 0.0195$ as Figure \ref{Unstable1-Applet} illustrates, then the equilibrium shifts to the stable fixed point $\frac{1}{3}$. Hence, the behavior of a dynamical system $f$ also depends on the initial point $x_0$. 
 \section{Behaviour of Wolbachia infected populations via discrete-time dynamical systems model}\label{section:dynamical}

In this section, we introduce our main object of study, the discrete-time dynamical system of the Wolbachia-infected population, (namely the bisexual population infected with a strain of Wolbachia). The population consists of four types of individuals, males without Wolbachia infection: $XY$, males with Wolbachia infection: $XY^{+}$, females without Wolbachia infection: $XX$, and females with Wolbachia infection: $XX^{+}$.

Let $x_0$ denote the frequency of the initial female population infected with Wolbachia within the total female population. Then $f(x_0)$ is the frequency of the infected female population of the offspring over the total female population. In a similar manner, $y_0$ denotes the frequency of the initial male population infected with Wolbachia within the total male population. Assume $p$ is the probability of the transmission of the Wolbachia infection from an individual onto its offspring \emph{(Cytoplasmic incompatibility (CI))}.  Therefore, $f: [0,1] \rightarrow [0,1]$ is a function whose input is the frequency of the Wolbachia-infected female population, and the output is the frequency of the Wolbachia-infected female population in the offspring. With this terminology, denote $f_n(x):=f(f(f...(f(x)))) = f^n(x)$ as the frequency of Wolbachia-infected female population in the $n^{th}$ generation when the initial infected female population frequency is $x$. 
 Let $g: \{ XX, XX^{+},  XY, XY^{+}\} \rightarrow [0,1]$ denote the frequency function of the phenotypes, and $g_{n}$ denote the frequency of the input phenotype in the $n^{th}$ generation. Now, analyze the initial generation with phenotype frequencies    
$$g(XX^{+})=x_{0},~~g(XX)=1-x_{0},~~g(XY^{+})=y_{0},~~g(XY)=1-y_{0}.$$ This population produces a gene pool with alleles $X$, $X^+$, $Y$, $Y^+$. Denote the frequency function of the alleles with $h$, that is $h_{n}^{(f)}$ denotes the allele frequency of the input allele within the female gene pool of the $n^{th}$ generation and $h_{n}^{(m)}$ denotes the allele frequency of the input allele within the male gene pool of the $n^{th}$ generation. 

For example, in the initial generation, female gene pool have frequencies
$$h_{0}^{(f)}(X^{+})=px_{0} \mbox{ and } h_{0}^{(f)}(X)=(1-p)x_{0}+1(1-x_{0})=1-px_{0},$$ whereas the male gene pool have frequencies $$h_{0}^{(m)}(X^{+})=\frac{1}{2}py_{0}=h_{0}^{(m)}(Y^{+}) \mbox{ and } h_{0}^{(m)}(X)=\frac{1}{2}(1-p)x_{0}+\frac{1}{2}(1-x_{0})=\frac{1}{2} (1-px_{0})=h_{0}^{(m)}(Y).$$ Consecutively, the first generation female gene pool has $h_{1}^{(f)}(X^{+})=px_{1}$ and the first generation male gene pool has $h_{1}^{(m)}(X^{+})=\dfrac{1}{2}py_{1}$ where $g_1(XX^{+})=x_{1}$ and $g_1(XY^{+})=y_{1}$. Iteratively, the $n^{th}$ generation gene pool frequencies become $h_{n}^{(f)}(X^{+})=px_{n}$ and $h_{0}^{(m)}(X^{+})=\dfrac{1}{2}py_{n}$ for any nonnegative integer $n$. The crossing table of the alleles in terms of the frequencies is given in Table \ref{fig:Allele-Table}.
\begin{table}[H]
\scalebox{0.85}{
\begin{tabular}{|c|c|c|c|c|}
\hline
\multicolumn{1}{|c|}{{\small Crossing}} & $X$ & $X^{+}$ & $Y$ & $Y^{+}$ \\ \hline
\multicolumn{1}{|c|}{$%
\begin{array}{c}
\text{ } \\ 
\text{ }%
\end{array}%
X%
\begin{array}{c}
\text{ } \\ 
\text{ }%
\end{array}%
$} & $h_{0}^{(f)}(X)h_{0}^{(m)}(X)$ & $---$ & $h_{0}^{(f)}(X)h_{0}^{(m)}(Y)$
& $---$ \\ \hline
\multicolumn{1}{|c|}{$%
\begin{array}{c}
\text{ } \\ 
\text{ }%
\end{array}%
X^{+}%
\begin{array}{c}
\text{ } \\ 
\text{ }%
\end{array}%
$} & $h_{0}^{(f)}(X^{+})h_{0}^{(m)}(X)$ & $%
h_{0}^{(f)}(X^{+})h_{0}^{(m)}(X^{+})$ & $h_{0}^{(f)}(X^{+})h_{0}^{(m)}(Y)$ & 
$h_{0}^{(f)}(X^{+})h_{0}^{(m)}(Y^{+})$ \\ \hline
\end{tabular}}

\bigskip 
\scalebox{0.85}{
\begin{tabular}{|c|c|c|c|c|}
\hline
{\small Crossing} & $X$ & $X^{+}$ & $Y$ & $Y^{+}$ \\ \hline
$%
\begin{array}{c}
\text{ } \\ 
\text{ }
\end{array}
X
\begin{array}{c}
\text{ } \\ 
\text{ }
\end{array}
$ & $(1-px_{0})\frac{1}{2}(1-py_{0})$ & $---$ & $(1-px_{0})\frac{1}{2}%
(1-py_{0})$ & $---$ \\ \hline
$
\begin{array}{c}
\text{ } \\ 
\text{ }
\end{array}
X^{+}
\begin{array}{c}
\text{ } \\ 
\text{ }
\end{array}
$ & $px_{0}\frac{1}{2}(1-py_{0})$ & $px_{0}\frac{1}{2}py_{0}$ & $px_{0}
\frac{1}{2}(1-py_{0})$ & $px_{0}\frac{1}{2}py_{0}$ \\ \hline
\end{tabular}}
\caption{\small Phenotype frequency table derived from the crossing of the alleles.}
\label{fig:Allele-Table} 
\end{table}
As Table \ref{fig:Allele-Table} demonstrates the frequency of Wolbachia-infected female population in the first generation is given by $$\frac{px_{0}\frac{1}{2}(1-py_{0}) + px_{0}\frac{1}{2}py_{0}}{px_{0}\frac{1}{2}(1-py_{0}) + px_{0}\frac{1}{2}py_{0} + (1-px_{0})\frac{1}{2}(1-py_{0})} =\frac{px_{0}}{1-py_{0}+p^{2}x_{0}y_{0}}.$$ 
Therefore, $f: [0,1] \rightarrow [0,1]$ is a function whose input is the frequency of the Wolbachia-infected female population, and the output is the frequency of the Wolbachia-infected female population in the offspring. For ease of notation, we drop the subscript zeros. The function depends on both variables $x$ and $y$ when $x \neq y$.  
\begin{equation*}
f(x,y):=\begin{cases}
\frac{px}{1 -px+ (px)^2}, & \textrm{if}\;x = y,\\
\frac{px}{1-py+p^{2}xy}, & \textrm{if}\; x \neq y
\end{cases}
\end{equation*} 
The derived equation is the discrete dynamical system of the Wolbachia-infected population with initial conditions $x$ and $y$. 
(In \citep{Fine}, $x$ and $y$ are taken to be equal, which is denoted by $B_a$.)

Note that in a given population if $x$ and $y$ are not the same initially, the ratio of the Wolbachia-infected female population to the total female population will equal to the ratio of the Wolbachia-infected male population to the total male population in the first generation. 
From  Table \ref{fig:Allele-Table} again, the frequency of Wolbachia-infected male population in the first generation is  
$$f_1(y) = f(y)= \frac{px\frac{1}{2}(1-py) + px\frac{1}{2}py}{px\frac{1}{2}(1-py) + px\frac{1}{2}py + (1-px)\frac{1}{2}(1-py)} = \frac{px}{1-py+p^{2}xy},$$ 
which is exactly the frequency of Wolbachia-infected female population in the first generation to the total female population.  Thus, Proposition \ref{xequalsy} is proved. 
\begin{prop}\label{xequalsy} For any initial Wolbachia-infected female frequency $x$ and initial Wolbachia-infected male frequency $y$, 
$f(x)=f(y)$, i.e. $g_1(XX^+)=f(x)=f(y)=g_1(XY^+)$. 
\end{prop}
Now, to understand the long behavior of the population, we can safely take the initial male/female Wolbachia-infected frequencies the same in the discrete dynamical system. So along the sequel, assume, $x=y$. Thus, the non-linear recurrence relation of the system is
\begin{equation}\label{recrel}
f_{n+1}(x)= \frac{p f_n(x)}{1 -p f_n(x) + (pf_n(x))^2}.
\end{equation}

When $p=0$, the Wolbachia infection is not transmitted vertically. 
Hence, we assume $p \neq 0$.   

If $x^*$ is a fixed point of $f$ then $x^*=f(x^*)= \frac{px^*}{1 -px^*+ (px^*)^2}$. 
Solving the quadratic equation for $x^*$, we get: 
$$ px^*= x^*(1 -px^*+ (px^*)^2) $$
$$ 0 = x^*((1-p) -px^*+ p^2(x^*)^2)$$ 
$$ x^* \in \left\{ 0, \frac{1+\sqrt{4p-3}}{2p}, \frac{1-\sqrt{4p-3}}{2p}\right\} $$
Depending on the value of $p$, the fixed point(s)  
\begin{equation*}
x^* =\begin{cases}
0, & \textrm{if}\;  0 < p < \frac{3}{4} .\\ 
0, 2/3 & \textrm{if}\; p = \frac{3}{4} ,\\
0,  \frac{1+\sqrt{4p-3}}{2p}, \frac{1-\sqrt{4p-3}}{2p} & \textrm{if}\; \frac{3}{4} < p < 1 ,\\
0, 1 & \textrm{if}\;  p = 1.\\
\end{cases}
\end{equation*} 

Now, we will analyze when these fixed points are stable or not analytically. 
Note that, $f^{\prime }(x^{\ast })=\frac{p(1-(px^{\ast })^{2})}{(1-px^{\ast
}+(px^{\ast })^{2})^{2}}$. By using Theorem \ref{thm:fixedpt} for each fixed point, the stable equilibrium points for different $p$ values are listed.  

\vspace{0.5 cm}
{\bf Case 1:}  if  $0 < p < \frac{3}{4}$, then $x^* = 0$ is a stable fixed point as $f'(0)=p < 1$.\\  

\begin{figure}[H]
\begin{center}
\subfigure[]{
\includegraphics[scale=0.3]{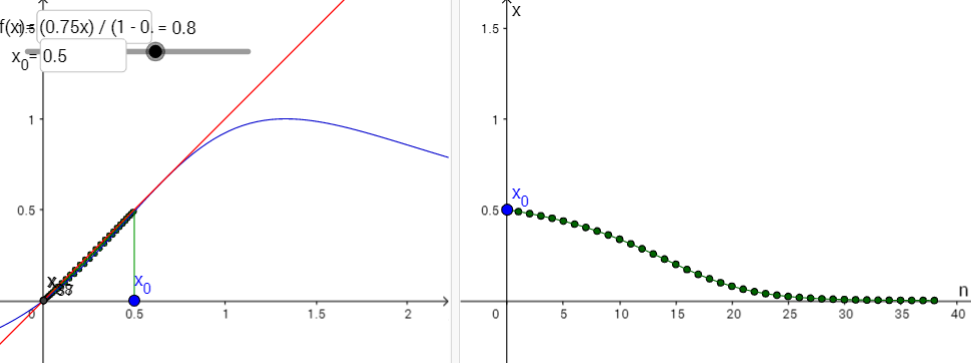}}
\subfigure[]{
\includegraphics[scale=0.3]{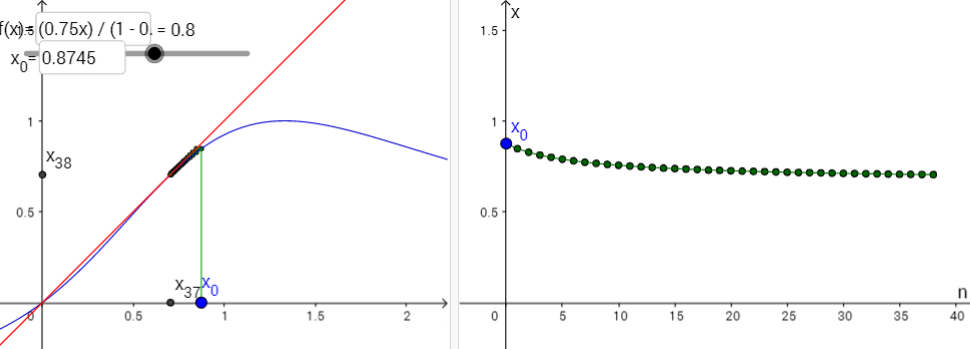}
}
\end{center}
\caption{Stability of fixed point $\frac{2}{3}$ for (a) $x_0=0.5$ and (b) $x_0=0.8745$. \\
{\it Produced by the applet: Visualizing function iteration via cobwebbing in \cite{MathInsight}} }
\label{fig:p=75} 
\end{figure}

{\bf Case 2:}  if  $p= \frac{3}{4}$,
\begin{equation*}
x^* =\begin{cases}
0 & \textrm{stable fixed point},\\  
\frac{2}{3} &  \textrm{stable / unstable fixed point},\\  
\end{cases}
\end{equation*}
When $p=0.75$, $f'(\frac{2}{3}) = 1$. Hence, we cannot determine whether $\frac{2}{3}$ is stable or not. For initial values less than  $\frac{2}{3}$, the fixed point $\frac{2}{3}$ is repellent. However, for initial values $x_0 > \frac{2}{3}$, $\frac{2}{3}$ behaves as an attracting fixed point. (See Figure \ref{fig:p=75}). 

\textbf{Case 3:} if $\frac{3}{4} < p < 1$: (In \cite{Fine}, if maternal and paternal CI values are equal, $d=w$ then the model gives $d \geq \frac{3}{4}$ which discusses this case only.) 

\[
x^{\ast }=\left\{ 
\begin{tabular}{cl}
$0$\medskip  & \ \ stable fixed point, \\ 
$\frac{1+\sqrt{4p-3}}{2p}$\medskip  & \ \ stable fixed point, \\ 
$\frac{1-\sqrt{4p-3}}{2p}$ & \ \ unstable fixed point,%
\end{tabular}%
\right. 
\]%
Again, $0$ is a stable fixed point, since $f^{\prime }(0)=p<1$. \newline
For $x^{\ast }=\frac{1+\sqrt{4p-3}}{2p},$ \ we have%
\[
px^{\ast }=\frac{1+\sqrt{4p-3}}{2}\ \text{\ and\ \ }(px^{\ast })^{2}=\left( 
\frac{1+\sqrt{4p-3}}{2}\right) ^{2}=p+\frac{1}{2}\sqrt{4p-3}-\frac{1}{2}.
\]
Then, $1-(px^{\ast })^{2}=\frac{3}{2}-\frac{1}{2}\sqrt{4p-3}-p$ \ and \ $%
1-px^{\ast }+(px^{\ast })^{2}=p.$\newline
Hence,%
\[
f^{\prime }(x^{\ast })=\frac{p(1-(px^{\ast })^{2})}{(1-px^{\ast }+(px^{\ast
})^{2})^{2}}=\frac{3-\sqrt{4p-3}}{2p}-1
\]%
On the other hand, $\frac{3}{4}<p<1$ implies $0<4p-3<1$ and so 
\[
-1<-\sqrt{4p-3}<4p-3<\sqrt{4p-3}< 1.
\]%
Now, the inequality $-1<-\sqrt{4p-3}<4p-3$ implies  $ 1 < \frac{1}{p}< \frac{3-\sqrt{4p-3}}{2p} < 2$, so \\ 
$$ 0 <  \frac{3-\sqrt{4p-3}}{2p}- 1 < 1.$$  Therefore, $|f^{\prime }(x^{\ast })|< 1$ and  $x^{\ast }=\frac{1+\sqrt{4p-3}}{2p
}$ is a stable fixed point.\newline
For $x^{\ast }=\frac{1-\sqrt{4p-3}}{2p},$ we have 
\[
f^{\prime }(x^{\ast })=\frac{p(1-(px^{\ast })^{2})}{(1-px^{\ast }+(px^{\ast
})^{2})^{2}}=\frac{3+\sqrt{4p-3}}{2p}-1
\]%
On the other hand, $\frac{3}{4}<p<1$ implies $0<4p-3<\sqrt{4p-3}<1$ and so $%
4p<3+\sqrt{4p-3}.$ Hence, 
\[ 2<\frac{3+\sqrt{
4p-3}}{2p}\ \mbox{ and so } 1<\frac{3+\sqrt{4p-3}}{2p}-1.\] Therefore, $f^{\prime
}(x^{\ast })>1$ and $x^{\ast }=\frac{1-\sqrt{4p-3}}{2p}$ is an unstable
fixed point. \\

{\bf Case 4:}  if  $p= 1$:

\begin{equation*}
x^* =\begin{cases}
0 & \textrm{stable/ unstable fixed point},\\  
1 &  \textrm{stable fixed point},\\  
\end{cases}
\end{equation*}



\section{Computations and data analysis}\label{section:Codes} 

Mathematical modeling of Wolbachia infection dynamics in a new host animal population highlights key parameters for the fate of Wolbachia infection in the animal community. According to the mathematical models, Wolbachia frequency has 3 different theoretical fates in a recently infected community, it may increase or decrease to a stable equilibrium point, or remain constant at an unstable genetic equilibrium point. The mathematical models suggest that cytoplasmic incompatibility induced by Wolbachia and initial Wolbachia-infected individuals introduced to the new community are key parameters dictating the fate of Wolbachia. 

As there are different Wolbachia evolutionary trajectories in a novel host community, differences in the interplay of the initial Wolbachia-infected individual frequency and Wolbachia cytoplasmic incompatibility intensity of Wolbachia may predict which evolutionary trajectory Wolbachia would pursue. For instance, in figures 6 and 7, mathematical models suggest that a Wolbachia strain infecting an insect community with a cytoplasmic incompatibility rate $> 0.95$ would either increase to an equilibrium point in the host population, almost reaching fixation at 0.997 or eradicate with a population frequency of 0. Mathematical models of Wolbachia infection of a community with a CI rate $> 0.95$ underlines that the final fate of Wolbachia in the community would depend on the initial frequency of Wolbachia infection in the community. Wolbachia with CI $> 0.95$, would increase to the stable equilibrium point of 0.997 reaching fixation if it has an initial frequency greater than 0.0555646363158. In the case of Wolbachia with CI $> 0.95$, with an increasing initial frequency above 0.0555646363158, there would be fewer generations required to reach the stable equilibrium point. However, if the initial frequency of Wolbachia with CI $> 0.95$ is less than 0.0555646363158, Wolbachia would be eradicated from the community meaning that it would reach a stable frequency equilibrium of 0. The number of generations for Wolbachia to become eradicated from the community would depend on the initial frequency of Wolbachia in the community. As the initial frequency of Wolbachia with CI $> 0.95$, decreases from 0.0555646363158 to 0 it would take fewer generations for Wolbachia to become eradicated. In a nutshell, Wolbachia shall have a very high cytoplasmic incompatibility rate and a significant initial frequency to reach an equilibrium point near fixation in the community. This may be expected from a sexual parasite which is detrimental to host fitness. It could only outcompete uninfected healthy hosts if it is readily available in the community and is effectively filtering out uninfected gametes from the community. Hence, mathematical models illustrate the implications of CI rate and initial Wolbachia frequency for the population dynamics of Wolbachia, emerging as a valuable tool to control pests and disease vectors.
 
 However, mathematical models may deviate from the phenomenon in nature. To test our model's predictive capacity, we simulated our model over iterations of generations with Wolbachia frequency and CI parameters similar to those in nature. Consequently, this part of the work will analyze the data from biology and interpretations of the computations from the papers \cite{isopod, Wolbachfigure}. The recurrence relation is iterated via a computer computation for 100 or 1000 generations depending on the speed of the population to attain the fixed value. The python program code used for computations is posted at the website https://github.com/canbluebird/Wolbachia. The computational results is compared with the experimental data found in literature. Note that $x$-axis represents the Wolbachia-infection frequency within the female population and $y$-axis represents the number of generations in Figures \ref{fig:a=b}, \ref{fig:p95unstablea-b}, \ref{fig:Can-p=75}, \ref{fig:p=70} \ref{fig:p=70anotb}. 

\subsection{Wolbachia in Terrestrial Isopod Populations }\label{subsection:Isopod} 

\begin{figure}[H]
\begin{center}
\includegraphics[scale=0.65]{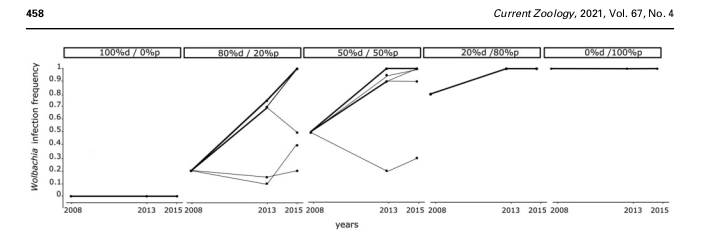}
\end{center}
\caption{An experimental data of terrestrial isopod from \citep{isopod}. Reprinted under the Creative Commons CC-BY-NC license.}
\label{fig:isopod} 
\end{figure}

\begin{figure}[H]
\begin{center}
\includegraphics[scale=0.66]{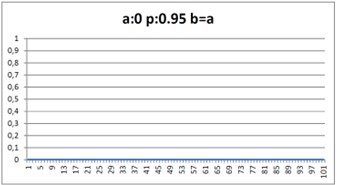}
\vspace*{3mm}
\includegraphics[scale=0.66]{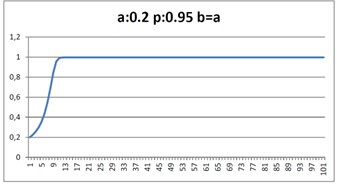} 
\includegraphics[scale=0.65]{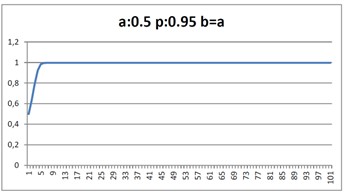} 
\vspace*{3mm}
\includegraphics[scale=0.64]{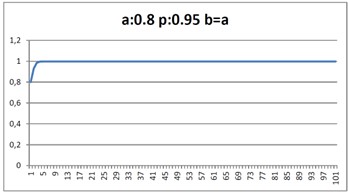}

\includegraphics[scale=0.65]{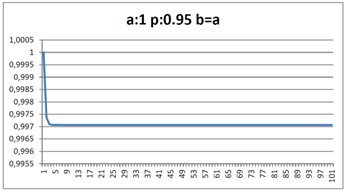}
\end{center}
\caption{The calculated data of the model for 100 generations.}
\label{fig:a=b} 
\end{figure}

In \citep{isopod}, the populations in each case, consist of a 1:1 ratio of male and female Wolbachia-infected terrestrial isopods. Also, the CI is considered to be greater than 95 \%. The Wolbachia-infected frequencies of the populations are measured within 8 years of the experiment. The frequency outcomes of the experiment is illustrated in Figure \ref{fig:isopod} which is taken from \citep{isopod}.    
In the simulations below $a:=x_0$ denote the frequency of the initial female population infected with Wolbachia
within the total female population and $b:=y_0$ denote the frequency of the initial male population infected with Wolbachia
within the total male population. 
To check the experimental data of \citep{isopod} (see, Figure \ref{fig:isopod}) with the mathematical model considered in Section \ref{section:dynamical}, proceed as follows:
\begin{itemize}
    \item 
Let $p=0.95$, $a=b$ where $a \in \{0, 0.2, 0.5, 0.8, 1 \}$ as in the terrestrial isopod experiment. 
    \item 
Calculate the Wolbachia-infected frequencies of the population for 100 generations by iterating the recurrence relation \ref{recrel} via the computer program.
    \item 
List the data in an excel file and graph the data (See Figure \ref{fig:a=b}). 
\end{itemize}

As expected from the behavior of the discrete-time dynamical system when $p=0.95$,
({\bf Case 3:} in Section \ref{section:dynamical}) , the stable fixed points are $0$ and $\displaystyle \frac{1+\sqrt{4p-3}}{2p}\sim 0.99706694$ and the unstable equilibrium is $\displaystyle \frac{1-\sqrt{4p-3}}{2p}\sim 0.0555646363158$. 
To illustrate the unstable equilibrium, the recurrence relation of the system (Equation \ref{recrel}) is computed for 1000 generations.
\newpage
\begin{figure}[H]
\begin{center}
\subfigure[]{
\includegraphics[scale=0.6]{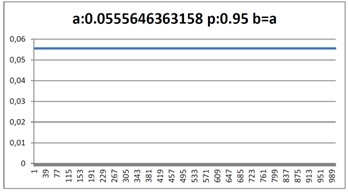}}
\subfigure[]{
\includegraphics[scale=0.6]{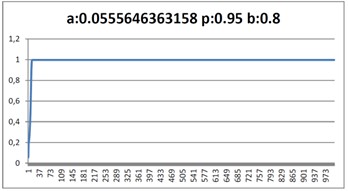}}
\subfigure[]{
\includegraphics[scale=0.6]{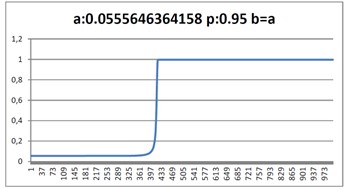}}
\subfigure[]{
\includegraphics[scale=0.6]{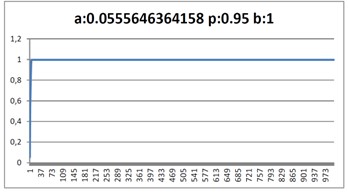}}
\subfigure[]{
\includegraphics[scale=0.6]{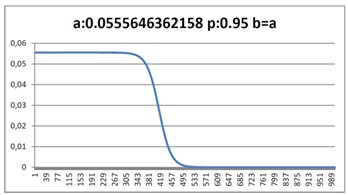}}
\subfigure[]{
\includegraphics[scale=0.6]{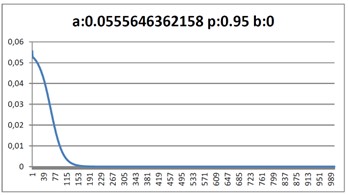}}
\end{center}
\caption{The stability of the model when $p=0.95$.}  
\label{fig:p95unstablea-b} 
\end{figure}
Recall that if $a\neq b$ initially, in the second generation the female Wolbachia-infected population to the total female population will be equal to the male Wolbachia-infected population frequency to the total male population which will be the Wolbachia infection frequency of the total population. Hence, in the discussion below, let us consider one parameter $a$ only. If 
$a=0$, the system stays at stable equilibrium $x^*=0$. 
If $0<a < \frac{1-\sqrt{4p-3}}{2p} \sim 0.0555646363158$ the system converges to the stable equilibrium $x^*=0 $ (Figure \ref{fig:p95unstablea-b} d,e). If $ a = \frac{1-\sqrt{4p-3}}{2p} \sim 0.0555646363158$, then the system is again at equilibrium and no change in the infected population occurs, see Figure \ref{fig:p95unstablea-b}(a). However, as $x^*=0.0555646363158$, is an unstable equilibrium, a slight change in the value of $a$ or $b$ towards one of two other stable equilibria points, results with a convergence of the system to that stable point, see Figure \ref{fig:p95unstablea-b}(b-f). 
On the other end, for values of $a >  \frac{1-\sqrt{4p-3}}{2p} \sim 0.0555646363158$  the system reaches a stable fixed point at $\displaystyle \frac{1+\sqrt{4p-3}}{2p}\sim 0.99706694$ when $p=0.95$ (Figure \ref{fig:p95unstablea-b} b,c). Also, note that within the dynamical system model constructed in this paper, this convergence may take up to 500 generations to occur, Figure \ref{fig:p95unstablea-b}(c,e).

Notice that when $a=b= \frac{1-\sqrt{4p-3}}{2p} \sim 0.0555646363158$, then the system is again at equilibrium and no change in the infected population occurs, see Figure \ref{fig:p95unstablea-b}(a). However, as $x^*=0.0555646363158$, is an unstable equilibrium, a slight change in the value of $a$ or $b$ towards one of two other stable equilibria points, results with a convergence of the system to that stable point, see Figure \ref{fig:p95unstablea-b}(b-f). Also, note that within the dynamical system model constructed in this paper, this convergence may take up to 500 generations to occur, Figure \ref{fig:p95unstablea-b}(c,e). 

\begin{figure}[H]
\begin{center}
\includegraphics[scale=0.35]{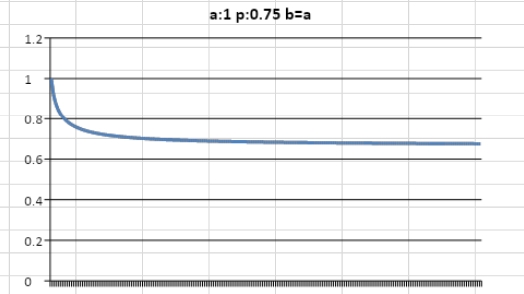}
\includegraphics[scale=0.35]{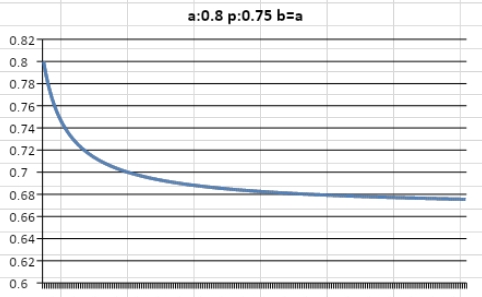}
\includegraphics[scale=0.35]{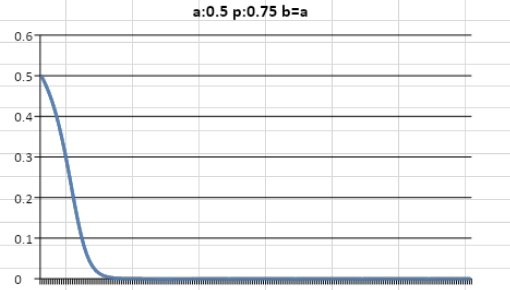}
\includegraphics[scale=0.35]{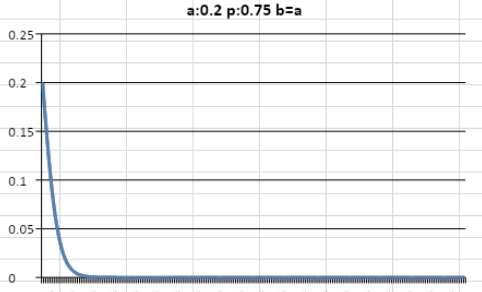}
\end{center}
\caption{The calculated data of the model for $p=0.75$.}
\label{fig:Can-p=75} 
\end{figure}

({\bf Case 2:} in Section \ref{section:dynamical})
However, when $p=0.75$, $\frac{1+\sqrt{4p-3}}{2p}=\frac{2}{3}$ and $f'(\frac{2}{3}) = 1$. Hence, we cannot determine whether $\frac{2}{3}$ is stable or not. For different initial values of the system, $\frac{2}{3}$ behaves as a converging / attracting fixed point, and some values when the point $\frac{2}{3}$ is repellent. When the initial value is $a=1$ or $0.8$, $\frac{2}{3}$ is an attracting fixed point. Whereas, if $a=0.5$ or $0.2$, then $\frac{2}{3}$ is a repellent fixed point and the system converges to the fixed point $x^*=0$ (See Figure \ref{fig:Can-p=75}).

\begin{figure}[H]
\begin{center}
\includegraphics[scale=0.5]{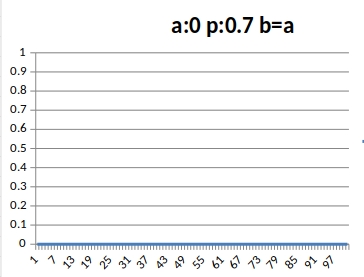}
\includegraphics[scale=0.5]{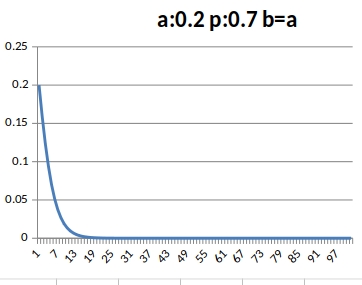}
\includegraphics[scale=0.5]{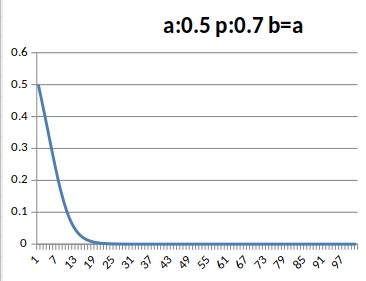}
\includegraphics[scale=0.5]{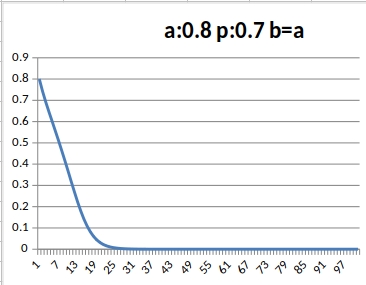}
\end{center}
\caption{$p=0.70$ when initial female/male infected frequencies are equal, i.e.  $a=b$}
\label{fig:p=70} 
\end{figure}

({\bf Case 1:} in Section \ref{section:dynamical})
Finally, when $p<0.75$, the only stable fixed point is $0$ as the calculations support. (See Figure \ref{fig:p=70}).
\newpage

\begin{figure}[H]
\begin{center}
\includegraphics[scale=0.45]{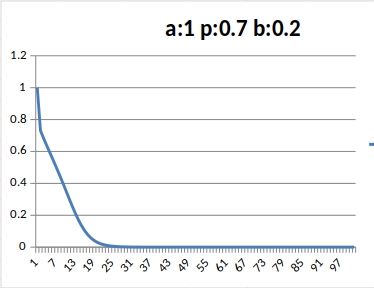}
\includegraphics[scale=0.45]{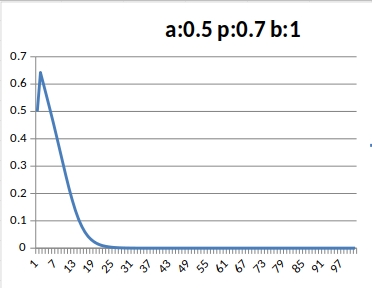}
\end{center}
\caption{The calculated data of the model for $p=0.70$ when $a \neq b$.}
\label{fig:p=70anotb}
\end{figure}

This is biologically interesting as it refers to: if the CI is less than 0.75 then no matter what the initial frequency of the Wolbachia infection within females (a) and/ or males (b) is, the system will have no infection in the long run. See that in Figure \ref{fig:p=70anotb}, both Wolbachia-infected male and female frequencies are non-zero, alas the system converges to zero. It might be interesting to support this mathematical model Equation \ref{recrel} with real data from a biological system, but we did not find any other study in the literature so far. Although, the paper \citep{isopod} mentions that CI is taken as 0.7 at some point, no experimental data is supplied.

\subsection{Wolbachia infection to control dengue disease in Mosquito Populations }\label{subsection:mosquito}
This subsection is devoted to the data and invasion model derived from the Wolbachia-infection in mosquito populations in \citep{Wolbachfigure}.
The data in Figure \ref{fig:Wolbachia1} is published as \citep[Figure 1(B)]{Wolbachfigure}. In this paper, the invasion dynamics of Wolbachia-infection is modeled as follows: 
\begin{figure}[H]
\begin{center}
\includegraphics[scale=0.5]{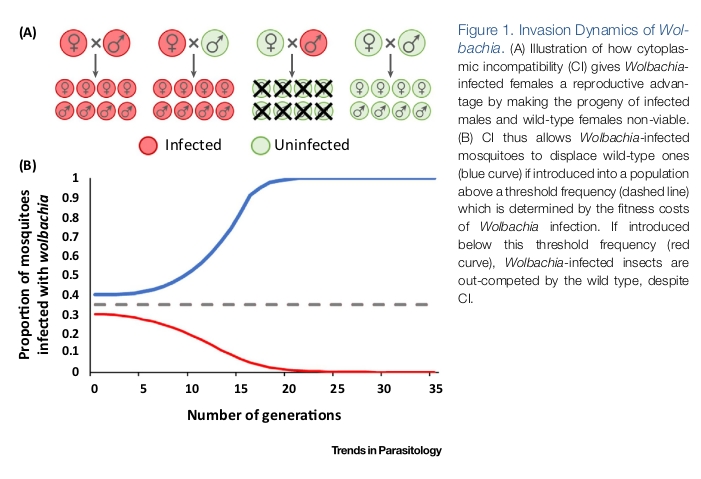}
\end{center}
\caption{Threshold value of Wolbachia-infection in mosquitoes graph from \citep[Figure 1(B)]{Wolbachfigure}.
Reprinted from Trends in Parasitology, 34(2), Ilaria Dorigatti, Clare McCormack, Gemma Nedjati-Gilani, Neil M. Ferguson. Using Wolbachia for Dengue Control: Insights from Modelling, 102-113, Copyright (2018), with permission from Elsevier.}
\label{fig:Wolbachia1}
\end{figure}
The paper discusses that the initial value of the dynamical system affects the equilibrium point. As to the behavior of the discrete-time dynamical system, if the initial infection frequency is above a threshold value then the system converges to approximately 1. How do we interpret this threshold in terms of the discussion of Equation \ref{recrel} in Section \ref{section:dynamical}? 
 
 This set-up is precisely an example of Case 3 in Section \ref{section:dynamical}. Notice that what the authors in \citep{Wolbachfigure} call the threshold value, 
 is the unstable equilibrium point of Case 3 in Section \ref{section:dynamical}. That is $ \frac{1-\sqrt{4p-3}}{2p} = 0.35$. Solving for the equation for $p$, we get $p \sim 0.79861$ which is Case 3: $\frac{3}{4} < p < 1$.  
The blue curve points out that the system reaches a stable fixed point at $ \frac{1+\sqrt{4p-3}}{2p} \sim 0.90216$ when the initial frequency of Wolbachia infection is 0.4. When the initial infection frequency is below the threshold value 0.35, the system converges to the stable fixed point zero, as the red curve states.

\subsection{Discussion}
Our simulated data with different CI rate, initial frequency, and female-to-male ratio parameters confirmed our mathematical models for population dynamics of Wolbachia infection. Nonetheless, simulations and mathematical models have to be validated with experimental data to confirm findings. We conclude the paper with a small discussion on how to set up an experiment to support the discrete dynamical system model achieved in this study of a Wolbachia-infected bisexual population with a fixed CI value. 

Rearing insect species with Wolbachia strains inducing CI with variable rates, and their wild-type counterparts, without Wolbachia infection would be required for validating mathematical suggestions. The widespread prevalence of CI-inducing Wolbachia infection in insect species flies renders a source for identifying Wolbachia-infected insect species with variable CI rates (Turelli et al 2022). Estimated stable population frequencies of Wolbachia strains in different species due to different CI rates, and the order of generation where the stable population frequency would be reached could be calculated and compared against Wolbachia data from generations of insect rearing. Insect populations could be mixed in different ratios of males and females, as well as infected and uninfected insects at the first generation to observe the impact of Wolbachia and gender frequency variations on where and when the stable equilibrium frequency of Wolbachia is reached in different insect populations. Nonetheless, the observed stable equilibrium frequencies of Wolbachia in experimental insect populations and when the stable frequency is reached may deviate in experimental observations from the suggestions of our mathematical models.

The mathematical models presented in this paper may present several sources for deviations in experimental observations. Insect population size may affect Wolbachia infection dynamics. The population size of insects in the mathematical models presented in this paper is reflective of infinite insect populations which is a challenge for experimental studies limited by insect population size. Moreover, models in this paper reflect discrete insect communities where there is no immigration and emigration, which could be recaptured in cage experiments but would lead to the divergence of experimental observations from the dynamics in nature. Furthermore, our model neglects the evolution of new Wolbachia mutations that may occur in experiments and confer host fitness benefits, as well as host mutations counteracting Wolbachia CI. Mutualistic host-Wolbachia evolution and evolutionary host response to Wolbachia infection are documented in nature (Turelli et al 2022). Therefore, deviations from mathematical models in observational Wolbachia frequencies may highlight Wolbachia and host evolution. To sum up, experimental insect populations may validate mathematical models, and illustrate the impact of host-Wolbachia evolution on Wolbachia frequencies. Comparing experimental data to observations from nature may underline the link between insect immigration and Wolbachia infection dynamics. In conclusion, host-Wolbachia infection and evolution dynamics emerge as a versatile source to derivate evolutionary mathematical models.

\end{document}